\documentclass[final,times,5p]{elsarticle}
\usepackage{ctable}
\usepackage{graphicx}
\usepackage{xspace}
\usepackage{hhline}
\usepackage{amsmath}
\usepackage{amsfonts}
\usepackage{hyperref}

\DeclareMathOperator{\Imag}{Im}
\newcommand{\ee}{\ensuremath{e^{+}e^{-}}\xspace}
\newcommand{\mumu}{\ensuremath{\mu^{+}\mu^{-}}}

\newcommand{\PP}{\ensuremath{\psi(2S)}\xspace}
\newcommand{\JP}{\ensuremath{J/\psi}\xspace}

\renewcommand{\epsilon}{\varepsilon}

\begin{document}
\title{Measurement of $R_{\text{uds}}$ and $R$ between 3.12 and 3.72 GeV at the KEDR detector}
\author[binp]{V.V.~Anashin}
\author[binp,nsu]{V.M.~Aulchenko}
\author[binp,nsu]{E.M.~Baldin} 
\author[binp]{A.K.~Barladyan}
\author[binp,nsu]{A.Yu.~Barnyakov}
\author[binp,nsu]{M.Yu.~Barnyakov}
\author[binp,nsu]{S.E.~Baru}
\author[binp]{I.Yu.~Basok}
\author[binp]{A.M.~Batrakov}
\author[binp,nsu]{A.E.~Blinov}
\author[binp,nsu,nstu]{V.E.~Blinov}
\author[binp,nsu]{A.V.~Bobrov}
\author[binp,nsu]{V.S.~Bobrovnikov}
\author[binp,nsu]{A.V.~Bogomyagkov}
\author[binp,nsu]{A.E.~Bondar}
\author[binp]{A.A.~Borodenko}
\author[binp,nsu]{A.R.~Buzykaev}
\author[binp,nsu]{S.I.~Eidelman}
\author[binp,nsu,nstu]{D.N.~Grigoriev}
\author[binp]{Yu.M.~Glukhovchenko}
\author[binp]{S.E.~Karnaev}
\author[binp]{G.V.~Karpov}
\author[binp]{S.V.~Karpov}
\author[binp]{P.V.~Kasyanenko}
\author[binp]{T.A.~Kharlamova}
\author[binp]{V.A.~Kiselev}
\author[binp]{V.V.~Kolmogorov}
\author[binp,nsu]{S.A.~Kononov}
\author[binp]{K.Yu.~Kotov}
\author[binp,nsu]{E.A.~Kravchenko}
\author[binp,nsu]{V.N.~Kudryavtsev}
\author[binp,nsu]{V.F.~Kulikov}
\author[binp,nstu]{G.Ya.~Kurkin}
\author[binp]{I.A.~Kuyanov}
\author[binp,nsu]{E.A.~Kuper}
\author[binp,nstu]{E.B.~Levichev}
\author[binp,nsu]{D.A.~Maksimov}
\author[binp]{V.M.~Malyshev}
\author[binp,nsu]{A.L.~Maslennikov}
\author[binp,nsu]{O.I.~Meshkov}
\author[binp]{S.I.~Mishnev}
\author[binp,nsu]{I.I.~Morozov}
\author[binp,nsu]{N.Yu.~Muchnoi}
\author[binp]{V.V.~Neufeld}
\author[binp]{S.A.~Nikitin}
\author[binp,nsu]{I.B.~Nikolaev}
\author[binp]{I.N.~Okunev}
\author[binp,nsu,nstu]{A.P.~Onuchin}
\author[binp]{S.B.~Oreshkin}
\author[binp,nsu]{A.A.~Osipov}
\author[binp,nstu]{I.V.~Ovtin}
\author[binp,nsu]{S.V.~Peleganchuk}
\author[binp,nstu]{S.G.~Pivovarov}
\author[binp]{P.A.~Piminov}
\author[binp]{V.V.~Petrov}
\author[binp,nsu]{V.G.~Prisekin}
\author[binp,nsu]{O.L.~Rezanova}
\author[binp,nsu]{A.A.~Ruban}
\author[binp]{V.K.~Sandyrev}
\author[binp]{G.A.~Savinov}
\author[binp,nsu]{A.G.~Shamov}
\author[binp]{D.N.~Shatilov}
\author[binp,nsu]{B.A.~Shwartz}
\author[binp]{E.A.~Simonov}
\author[binp]{S.V.~Sinyatkin}
\author[binp]{A.N.~Skrinsky}
\author[binp,nsu]{A.V.~Sokolov}
\author[binp,nsu]{A.M.~Sukharev}
\author[binp,nsu]{E.V.~Starostina}
\author[binp,nsu]{A.A.~Talyshev}
\author[binp,nsu]{V.A.~Tayursky}
\author[binp,nsu]{V.I.~Telnov}
\author[binp,nsu]{Yu.A.~Tikhonov}
\author[binp,nsu]{K.Yu.~Todyshev \corref{cor}}
\cortext[cor]{Corresponding author, e-mail: todyshev@inp.nsk.su}
\author[binp]{G.M.~Tumaikin}
\author[binp]{Yu.V.~Usov}
\author[binp]{A.I.~Vorobiov}
\author[binp,nsu]{V.N.~Zhilich}
\author[binp,nsu]{V.V.~Zhulanov}
\author[binp,nsu]{A.N.~Zhuravlev}
 \address[binp]{Budker Institute of Nuclear Physics, 11, akademika
 Lavrentieva prospect,  Novosibirsk, 630090, Russia}
 \address[nsu]{Novosibirsk State University, 2, Pirogova street,  Novosibirsk, 630090, Russia}
 \address[nstu]{Novosibirsk State Technical University, 20, Karl Marx
  prospect,  Novosibirsk, 630092, Russia}

\begin{abstract} 
Using the KEDR detector at the VEPP-4M \ee collider, we have measured
the values of $R_{\text{uds}}$ and $R$ at seven points  of the center-of-mass energy between 3.12 and 3.72 GeV. 
The total achieved accuracy  is about or better than  $3.3\%$ at most of energy points  with a systematic uncertainty of about $2.1\%$. At the moment it is the most accurate measurement of $R(s)$ in this energy
range.
\end{abstract}
\maketitle
\section{Introduction}\label{sec:intro}
The quantity $R$ is defined as the ratio of the radiatively corrected
total hadronic cross section in electron-positron annihilation 
to the lowest-order QED cross section  of the muon pair production. 
The precise $R(s)$ measurements are critical for determination 
of the value of the strong coupling constant $\alpha_s(s)$ and heavy 
quark masses~\cite{quark},
the anomalous magnetic moment of the muon $(g-2)_\mu$ and the value of the
electromagnetic fine structure constant at the $Z^0$ peak
$\alpha(M_Z^2)$~\cite{dhmz,hlmnt}.

Several  experiments contributed to  the  $R(s)$ measurement  
in the energy range between 3.12 and 3.72~GeV \cite{Mark1:R1977,PLUTO:R1977,GG2:R1979,MARK2:R1980,MARK1:R1982,Bai:1999pk,BES:R2002,BES:R2006,BES:R2009}. The precision of these 
measurements does not exceed $5\%$  for all experiments except  
BES-II~\cite{BES:R2009}, in which the accuracy of about 3.3\%  was reached 
at 3.07 and 3.65~GeV, but that is not enough for reliable calculation of 
the dispersion integrals in the whole energy range.
It should be noted that systematic
uncertainties dominate in all $R(s)$ measurements,
thus there is good motivation for new experiments on the precise
determination of $R(s)$ in this energy range, particularly important for 
$\alpha(M_Z^2)$.

In 2011 the region of the $J/\psi$ and $\psi(2S)$ resonances
was scanned in the KEDR \cite{KEDR:Det} experiment with an integrated
luminosity of  about 1.4~pb$^{-1}$. In the data analysis presented
below we tried to minimize correlations of systematic uncertainties with
those in similar experiments by BES. 

\section{VEPP-4M collider and KEDR detector}\label{sec:VEPP}
The  \ee collider VEPP-4M ~\cite{Anashin:1998sj}  was designed 
to operate in the wide range of the beam energy 1$\div$5.5 GeV 
in the 2$\times$2 bunches mode.
The peak luminosity of VEPP-4M is 
about~\(2\!\times\!10^{30}\,\text{cm}^{-2}\text{s}^{-1}\) in the vicinity of \PP.

The collider is well equipped for a precise beam energy determination.
The beam energy in dedicated calibration runs is measured
using the resonant depolarization
method (RDM)~\cite{Bukin:1975db,Skrinsky:1989ie} with
the relative accuracy of about \(10^{-6}\). 
The results of RDM calibrations can be interpolated to determine
the energy during data taking with the accuracy 
of about 10~keV~\cite{Aulchenko:2003qq,MASS::KEDR2015}.
Continuous energy monitoring is performed using 
the infrared light Compton backscattering~\cite{CBS} with 
the accuracy $\sim$ 60~keV. The Compton backscattering also
allows for the beam energy spread determination with the accuracy about 5\%.

The KEDR detector is described in Ref.~\cite{KEDR:Det}.
The detector consists of  the vertex detector (VD), 
drift chamber (DC),  time-of-flight (TOF) system of scintillation counters,
particle identification system based on the aerogel Cherenkov
counters, electromagnetic calorimeter (liquid krypton in the barrel part and
CsI crystals in the endcaps), superconducting solenoid 
and muon system inside the magnet yoke. The superconducting 
solenoid provides a longitudinal magnetic field of 0.6~T.
The detector is equipped with a tagging system of scattered electrons  
for two-photon studies.
The on-line luminosity measurement is provided by two independent
single bremsstrahlung monitors.

The trigger has two hardware levels:
the primary  (PT) and the secondary one (ST)~\cite{TALYSHEV}.
The PT operates using signals from the TOF counters and fast signals
from the CsI and LKr calorimeters,
whereas the ST uses optimally shaped calorimeter signals
and the information from VD, DC and the TOF system.

\section{Experiment}\label{sec:exp}
\begin{figure*}[t!]
\begin{center}
\includegraphics[width=0.8\textwidth]{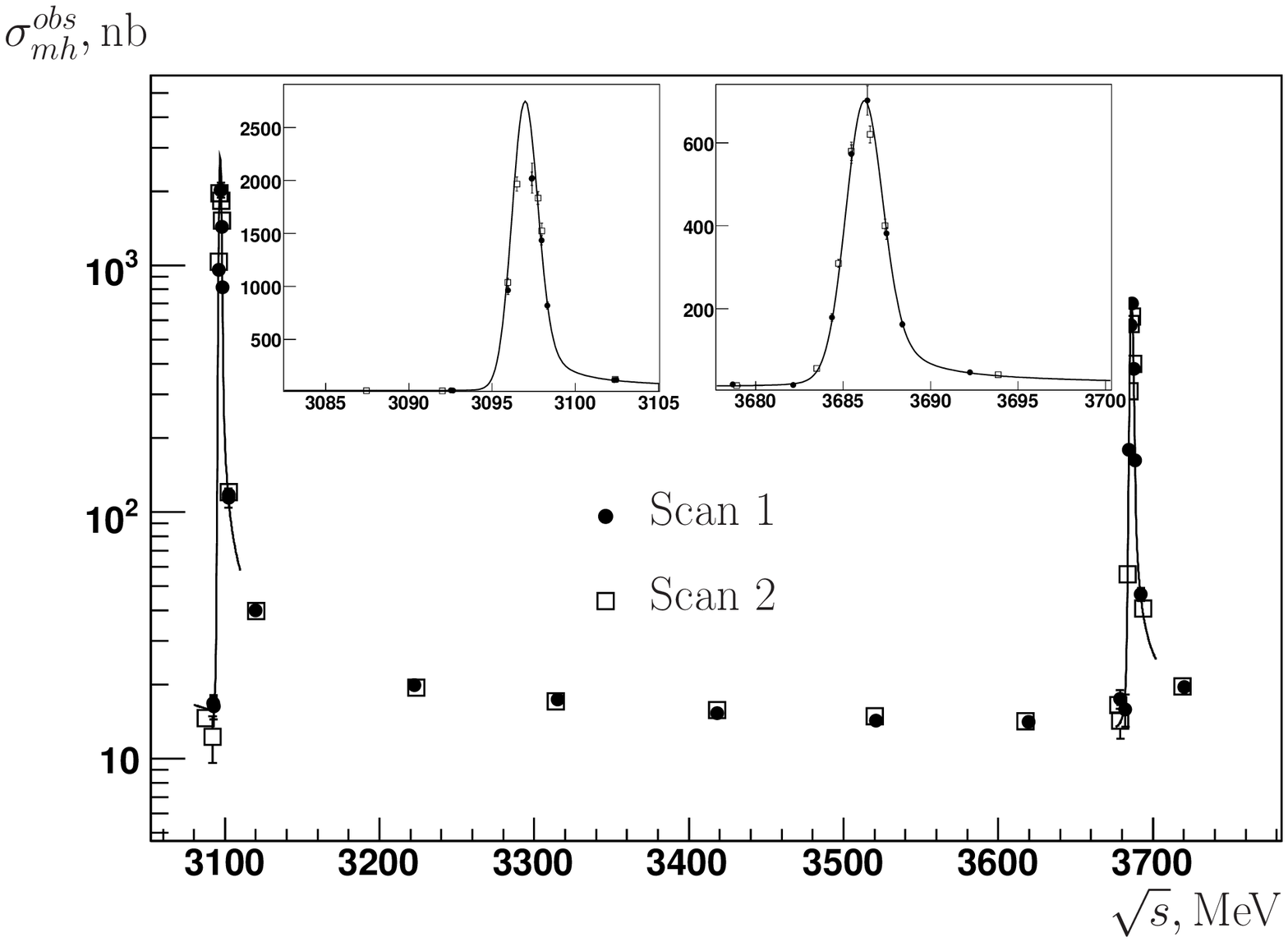}
\caption{{The observed multihadronic cross section as a function 
of the c.m. energy for the two scans. 
The curves are the result of the fits of the narrow resonances.
The inserts show closeup of the $J/\psi$ and $\psi(2S)$ regions.
\label{scans_picture}
}}
\end{center}
\end{figure*}

The goal of the experiment was a measurement of the total hadronic
cross section at seven equidistant points between 3.12 and 3.72 GeV.
Two scans of the region were performed. The actual energies
determined using the Compton backscattering method and the integrated
luminosity at the points are presented in Table~\ref{tab:epoints}.
The table also presents the relative contributions of the \JP and \PP 
in the observed cross section dominated by their radiative tails.
To determine them without  external data, the additional
data samples of about 0.4~pb$^{-1}$ were collected at ten points in the peak
regions. The data points and the resonance fits are shown in 
Fig.~\ref{scans_picture}.

\renewcommand{\arraystretch}{1.7}
\setlength{\tabcolsep}{3pt}
\begin{center}
\begin{table}[h!]
\caption{{\label{tab:epoints} Center-of-mass energy $\sqrt{s}$,  integrated luminosity $\int\!\!\mathcal{L}dt$ and relative contribution 
of the $J/\psi$ and $\psi(2S)$ resonances to the  observed multihadronic cross section.}} 
\begin{tabular}{lcccc}
\specialrule{.14em}{.07em}{.07em} 
\firsthline
Point & $\sqrt{s}$, MeV &  $\int\!\!\mathcal{L}dt$, nb$^{-1}$ & $\frac{\sigma_{J/\psi}}{\sigma_{\text{obs}}}$,$\%$ &$\frac{\sigma_{\psi(2S)}}{\sigma_{\text{obs}}}$,$\%$   \\  \hline
 \multicolumn{5}{c}{Scan 1} \\ \hline
1& $3119.8\pm 0.2$  &  ~~$64.31 \pm 0.72$ & 59.6 &\\ \hline
2& $3222.4\pm 0.2$  &  ~~$74.79 \pm 0.80$ & 22.9 &\\ \hline
3& $3315.2\pm 0.2$  &  ~~$83.25 \pm 0.87$ & 14.8 &\\ \hline
4& $3418.1\pm 0.2$  &  ~~$95.68 \pm 0.97$ & 10.9 &\\ \hline
5& $3521.0\pm 0.2$  &    $112.36\pm 1.08$ & 8.3  &\\ \hline    
6& $3619.7\pm 0.2$  &  ~~$34.72 \pm 0.61$ & 5.6 & \\ \hline
7& $3720.4\pm 0.2$  &  ~~$55.57 \pm 0.80$ & 3.6 & 29.7\\ \hline   
  \multicolumn{5}{c}{Scan 2}      \\ \hline        
1&$3120.1  \pm 0.2$ &  $54.46 \pm 0.63$  & 58.3 & \\ \hline
2&$3223.6  \pm 0.2$ &  $65.77 \pm 0.88$  & 23.0 & \\ \hline
3&$3313.9  \pm 0.2$ &  $50.93 \pm 0.61$  & 14.9 & \\ \hline
4&$3418.4  \pm 0.2$ &  $66.88 \pm 0.88$  & 10.4 & \\ \hline
5&$3520.3  \pm 0.2$ &  $59.33 \pm 0.67$  & 7.9   &  \\ \hline    
6&$3617.6  \pm 0.2$ &  $83.35 \pm 0.95$  & 5.6    &  \\ \hline
7&$3718.9  \pm 0.2$ &  $103.66 \pm 1.05 $  & 3.5 &  30.5 \\      
\hhline{=====}
\end{tabular} 
\end{table}
\end{center}

\renewcommand{\arraystretch}{1.}

The energies of the points in two scans are not the same 
because of the inaccuracy of the collider energy setting,
but they are close enough to allow for summation of data samples.

\section{Data analysis}\label{sec:data}
\subsection{Analysis procedure}\label{subsec:proc}
The observed hadronic annihilation cross section 
was determined from
\begin{equation}
\sigma_{\text{obs}}(s)=\frac{N_{\text{mh}} - N_{\text{res.bg.}}}{\int\!\!\mathcal{L}dt},
\label{eq:sigmaobs}
\end{equation}
where $N_{\text{mh}}$ is the number of events that pass hadronic
selection criteria,
$N_{\text{res.bg.}}$ is the residual machine background evaluated as
 discussed in Sec.~\ref{sec:background},
and $\int\!\!\mathcal{L}dt $ is the integrated luminosity.

For the given observed cross section the $R$ value was calculated
as follows:
\begin{equation}
\label{eq:R}
R = \frac{\sigma_{\text{obs}}(s)-
\sum\epsilon_{\text{bg}}(s)\,\sigma_{\text{bg}}(s) -
\sum\epsilon_{\psi}(s)\,\sigma_{\psi}(s)
 } {\epsilon(s)\, (1+\delta(s)) \,\sigma_{\mu\mu}^0(s)},
\end{equation}
where $\sigma_{\mu\mu}^0(s)$ is the Born cross section
for $\ee\!\to\!\mumu$ and
$\epsilon(s)$ is the detection efficiency for the single photon annihilation
to hadrons. The second term in the numerator corresponds to the physical
background from  $e^+e^-$, $\mu^+\mu^-$ production, $\tau^+\tau^-$
production above threshold and two-photon processes.
The third term represents a contribution of the \JP and \PP.
Unlike Refs.~\cite{Bai:1999pk,BES:R2002,BES:R2006,BES:R2009}, we considered them 
explicitly instead of including in the radiation correction $\delta(s)$.

The detection efficiencies $\epsilon$ and $\epsilon_{\text{bg}}$ were
determined from simulation. The efficiencies
$\epsilon_{\psi}$ were found by fitting the resonance regions.
The resonances were fitted separately in each scan, 
the free parameters were the detection efficiency at the world average
values of the leptonic width $\Gamma_{ee}$ and its product by the hadronic
branching fraction $\mathcal{B}_{h}$,
the machine energy spread 
and the magnitude of the continuum cross section observed at the reference 
point below the resonance. 
Calculations of a narrow resonance cross section  and fits
are described in more detail in Refs.~\cite{MASS::KEDR2015,psi2S:2012}.
The $J/\psi$ and $\psi(2S)$ detection efficiencies
obtained and the $\chi^2$ probabilities of the fits 
are presented in Table~\ref{tab:fits}. The fitted values of the collision
energy spread are also presented. They are not the same for the two scans
because of variation of the accelerator regime, however, the energy spread
is stable during a few days of operation in the resonant regions. 
The quoted values agree with the results of the energy spread
determination using Compton backscattering within the accuracy provided by 
this method.
\renewcommand{\arraystretch}{1.2}
\setlength{\tabcolsep}{3pt}
\begin{table*}[t!]
\begin{center}
\caption{ \label{tab:fits} { Efficiency, energy spread and $\chi^2$ 
probability  of the fits of the $J/\psi$ and $\psi(2S)$ resonances
(statistical errors only are presented).}}
\begin{tabular}{lcccccc} 
\specialrule{.14em}{.07em}{.07em} 
\firsthline
               &  $\epsilon_{J/\psi}$& $\sigma_W(J/\psi)$, MeV &$P(\chi^2)$, $\%$ & $\epsilon_{\psi(2S)}$ & $\sigma_W(\psi(2S))$, MeV & $P(\chi^2)$, $\%$ \\\hline
Scan 1         &  $0.760 \pm 0.013$ & $0.741\pm 0.005$ &77.6  &   $0.838 \pm  0.023$& $0.961 \pm  0.033$ & 44.9  \\ \hline
Scan 2         &  $0.751 \pm 0.014$ & $0.761\pm 0.007$ &18.5  &   $0.830 \pm  0.020$& $1.049 \pm  0.054$ & 73.3  \\
 \hhline{=======}
\end{tabular}
\end{center}
\end{table*}
\renewcommand{\arraystretch}{1.}

It should be noted that the tail cross section
$\epsilon_{\psi}(s)\,\sigma_{\psi}(s)$ 
depends on the $\epsilon_{\psi}\Gamma_{ee}\mathcal{B}_{h}$ product
and thus is not sensitive to the world average values 
of the leptonic width $\Gamma_{ee}$ and the hadronic branching fraction $\mathcal{B}_{h}$ employed.

In our approach the radiative correction factor can be written as
\begin{equation}
\label{eq:RadDelta}
1+\delta(s)=\int\!\frac{dx}{1\!-\!x}\, 
\frac{\mathcal{F}(s,x)}{\big|1-\tilde{\Pi}((1\!-\!x)s)\big|^2}\,
\frac{\tilde{R}((1\!-\!x)s)\,\epsilon((1\!-\!x)s)}{R(s)\,\epsilon(s)},
\end{equation}
where $\mathcal{F}(s,x)$ --  the radiative correction kernel~\cite{KF:1985}.
The variable $x$ is a fraction of $s$ lost due to initial state
radiation.
The vacuum polarization operator $\tilde{\Pi}$ and  the quantity $\tilde{R}$  do not include a contribution of the $J/\psi$ and $\psi(2S)$ resonances, details of the calculation are presented in Section~\ref{sec:radeff}.

It should be noted that in the approach described above we obtain the $R_{\text{uds}}$ value.
To get the quantity $R$, it is necessary to add the contribution of narrow resonances.
In the following we shall use $R$ instead of $R_{\text{uds}}$ until Section~\ref{sec:errsummary}.

\subsection{Monte Carlo simulation}\label{sec:MC}
Simulation of the experiment was performed
in the frame of the  GEANT package, version 3.21~\cite{GEANT:Tool}.

Single-photon annihilation to hadrons below $D\overline{D}$ threshold
($\text{uds}$ continuum) was simulated
using the JETSET~7.4 code \cite{JETSET,PYTHIA64} with the parameters tuned 
at each energy point.
As an alternative, we employed the LUARLW \cite{LUARLW:2001}
generator which was kindly provided by the BES collaboration. 

\begin{center}
\begin{figure*}[!ht]
\begin{center}
\includegraphics[width=0.346\textwidth,height=0.22\textheight]{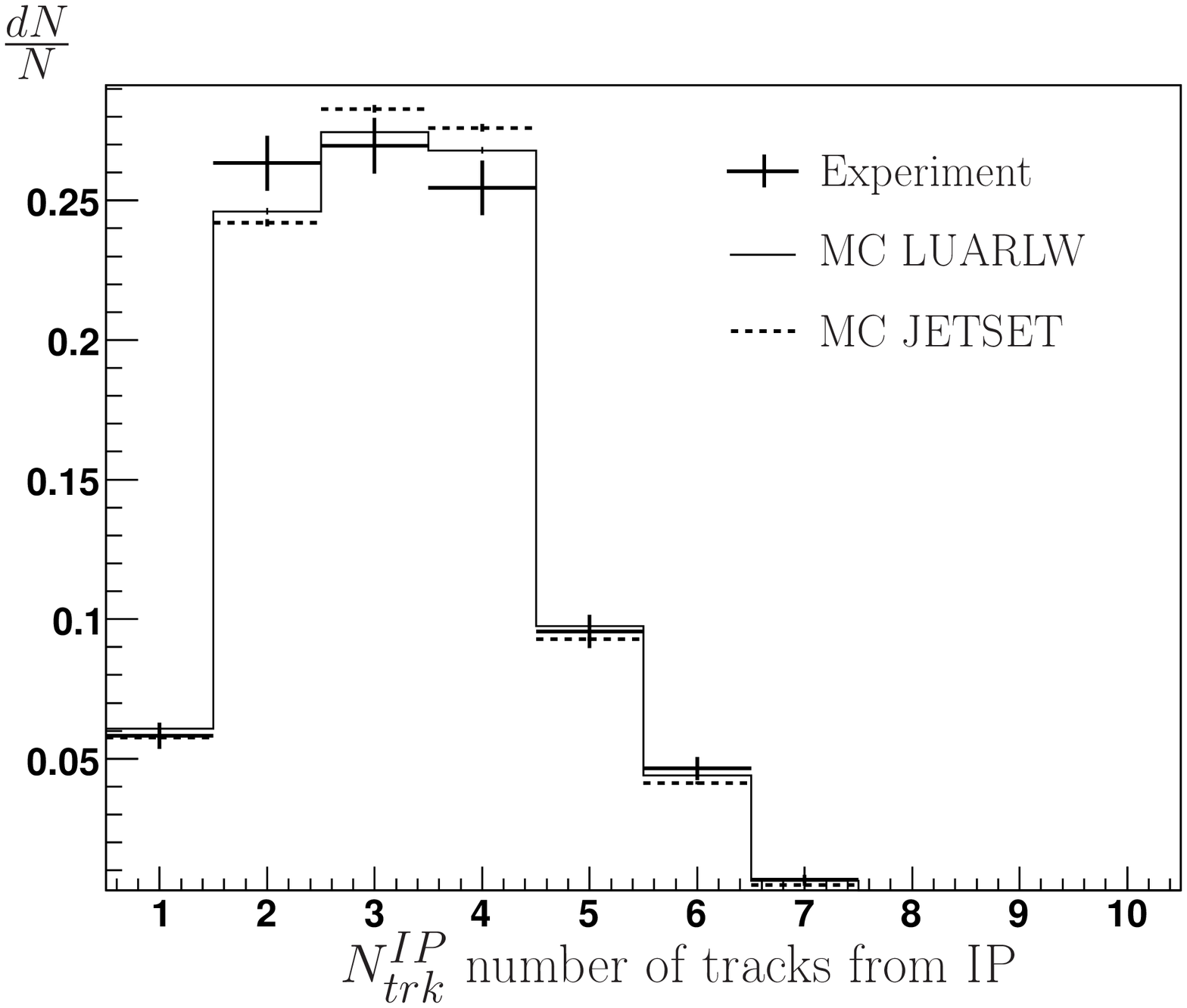}
\includegraphics[width=0.351\textwidth,height=0.22\textheight]{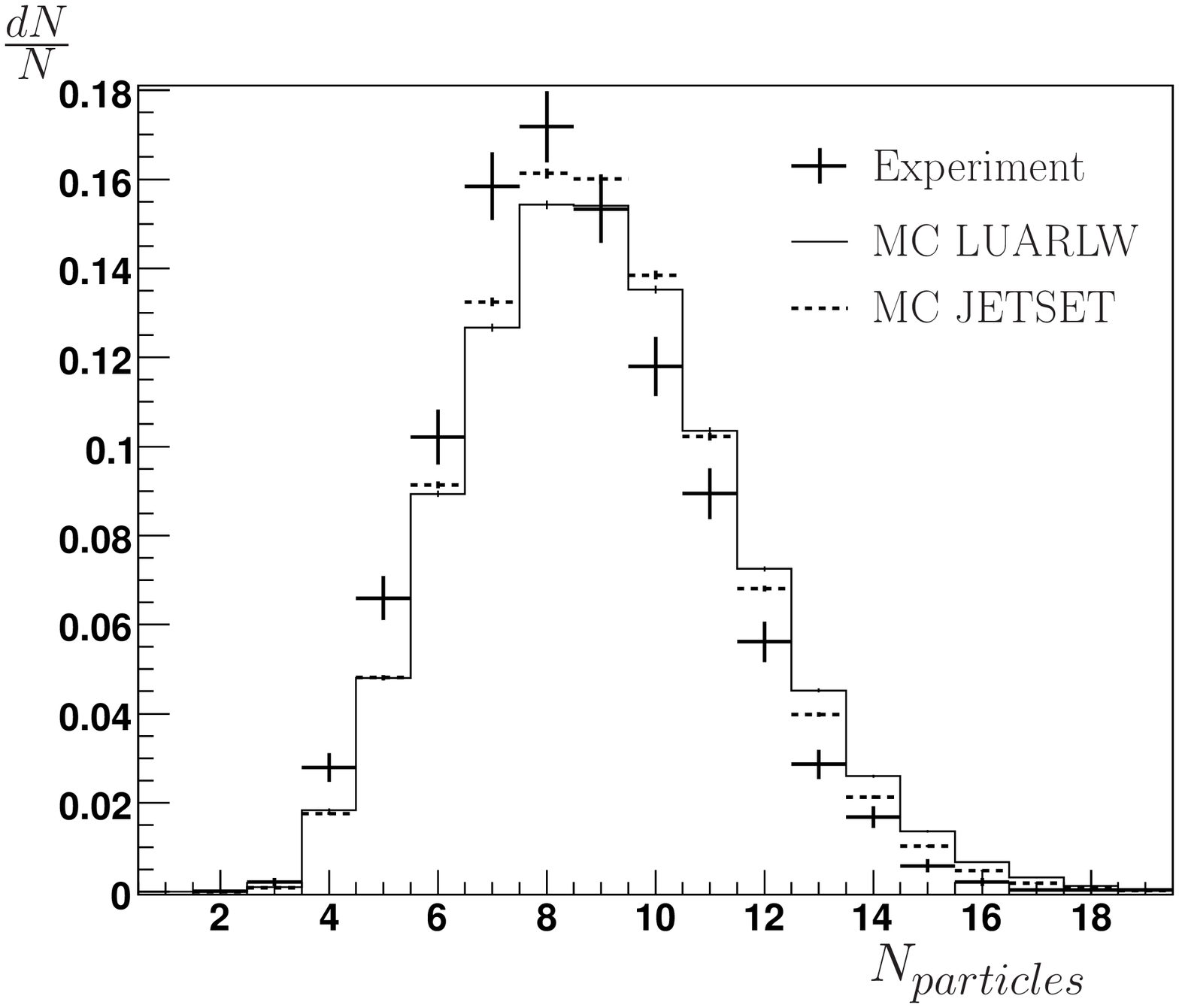} \\
\includegraphics[width=0.35\textwidth,height=0.22\textheight]{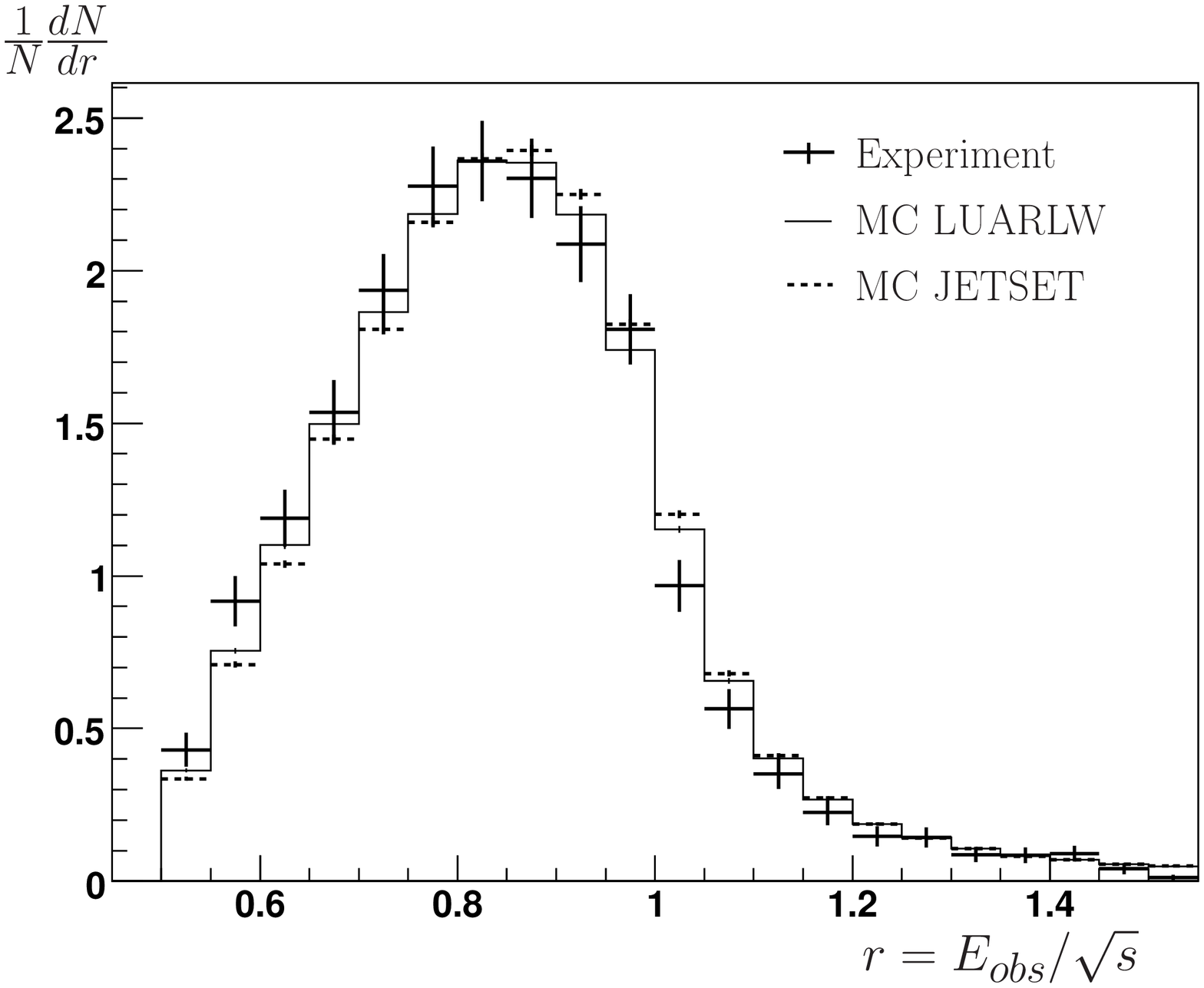}
\includegraphics[width=0.3525\textwidth,height=0.216\textheight]{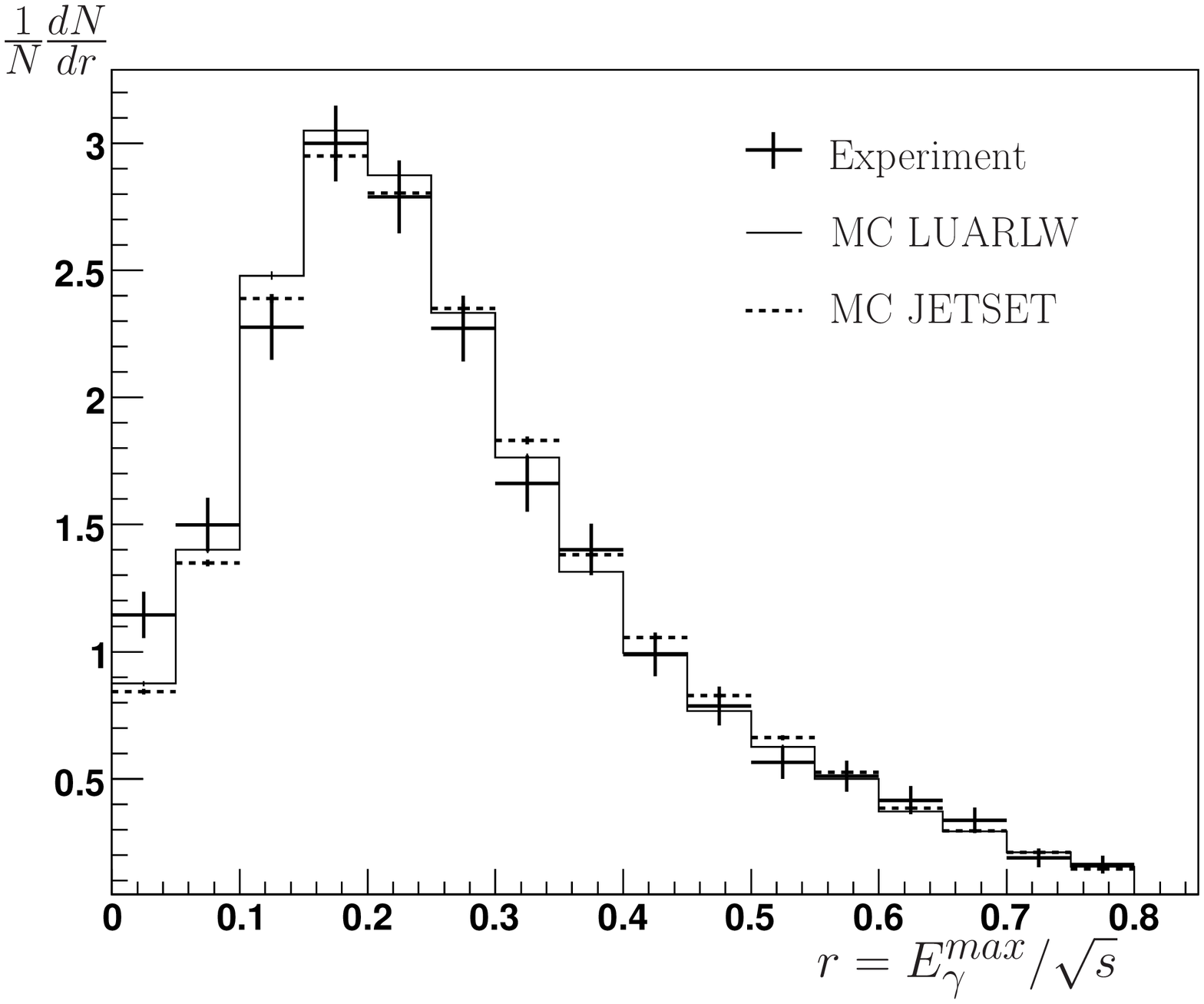} \\
\includegraphics[width=0.353\textwidth,height=0.22\textheight]{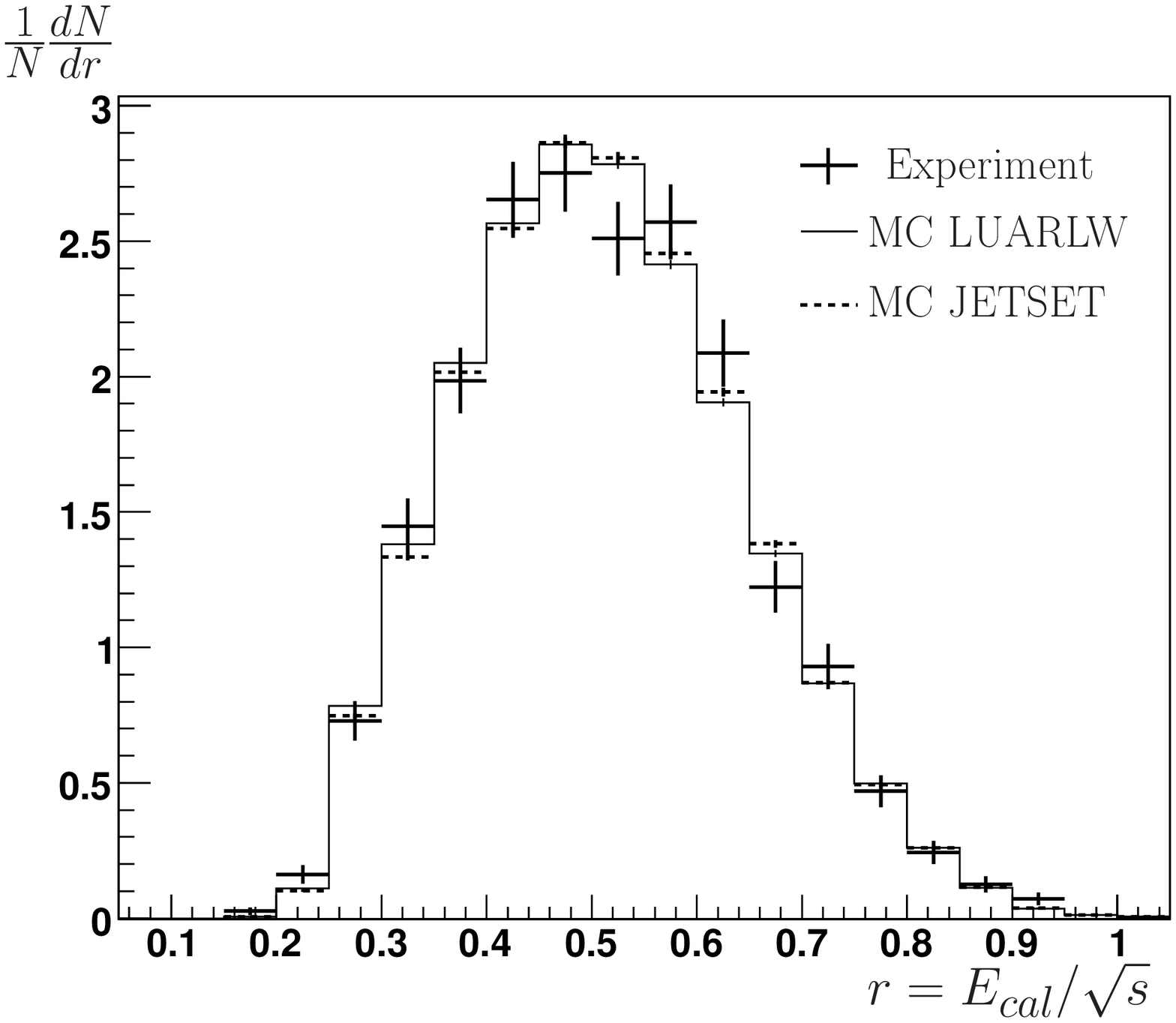}
\includegraphics[width=0.3495\textwidth,height=0.221\textheight]{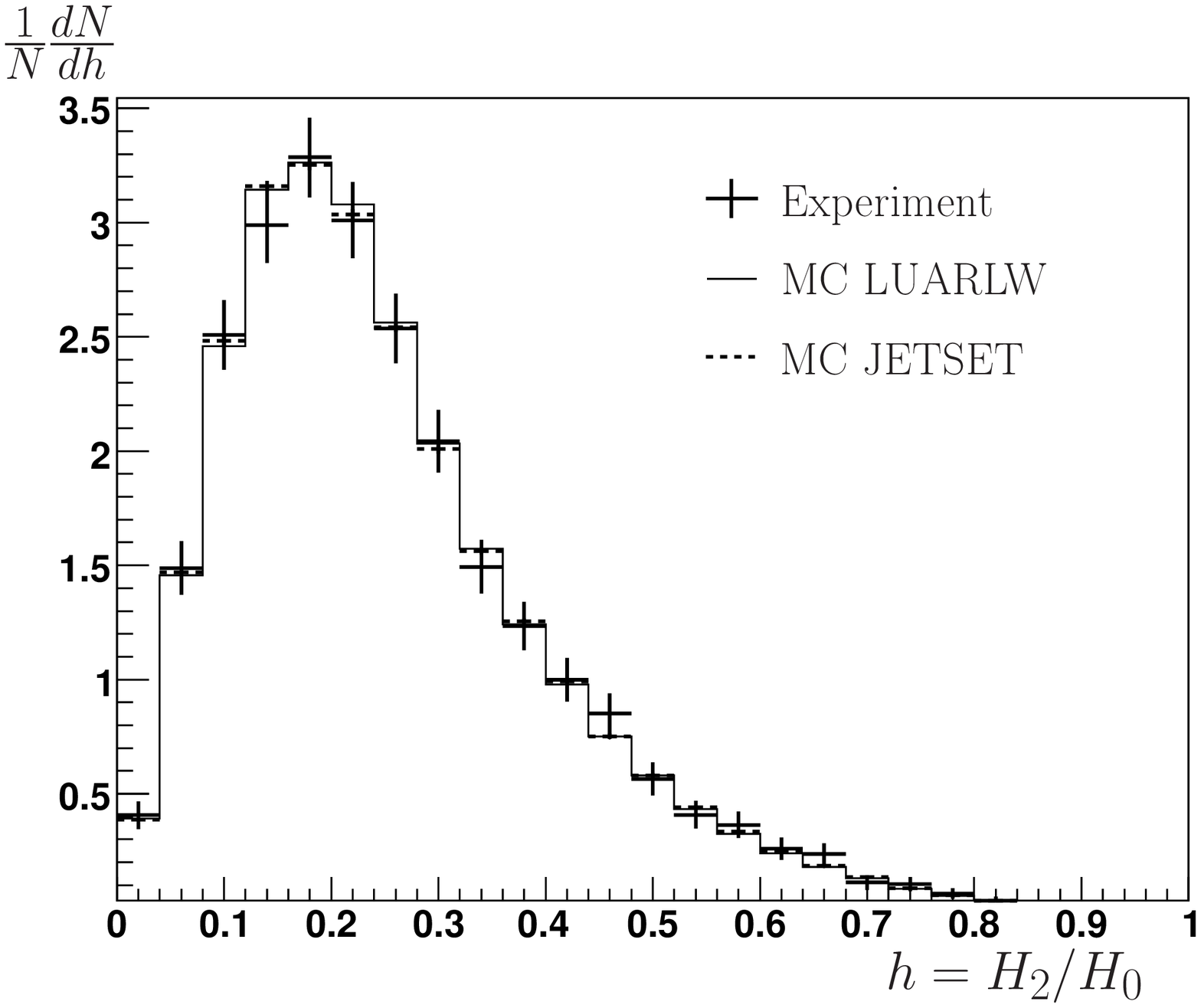} \\
\includegraphics[width=0.352\textwidth,height=0.231\textheight]{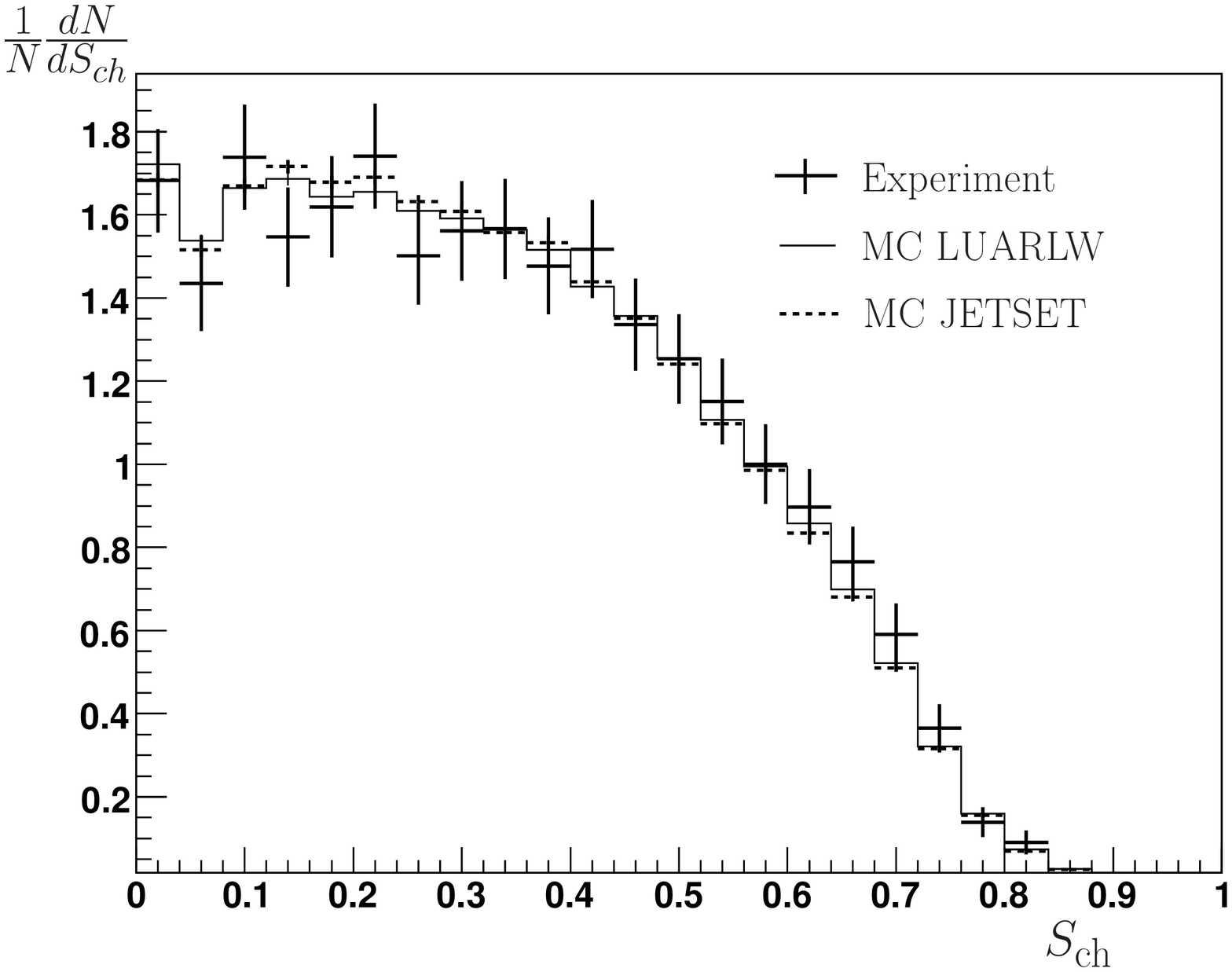}  
\includegraphics[width=0.354\textwidth,height=0.2343\textheight]{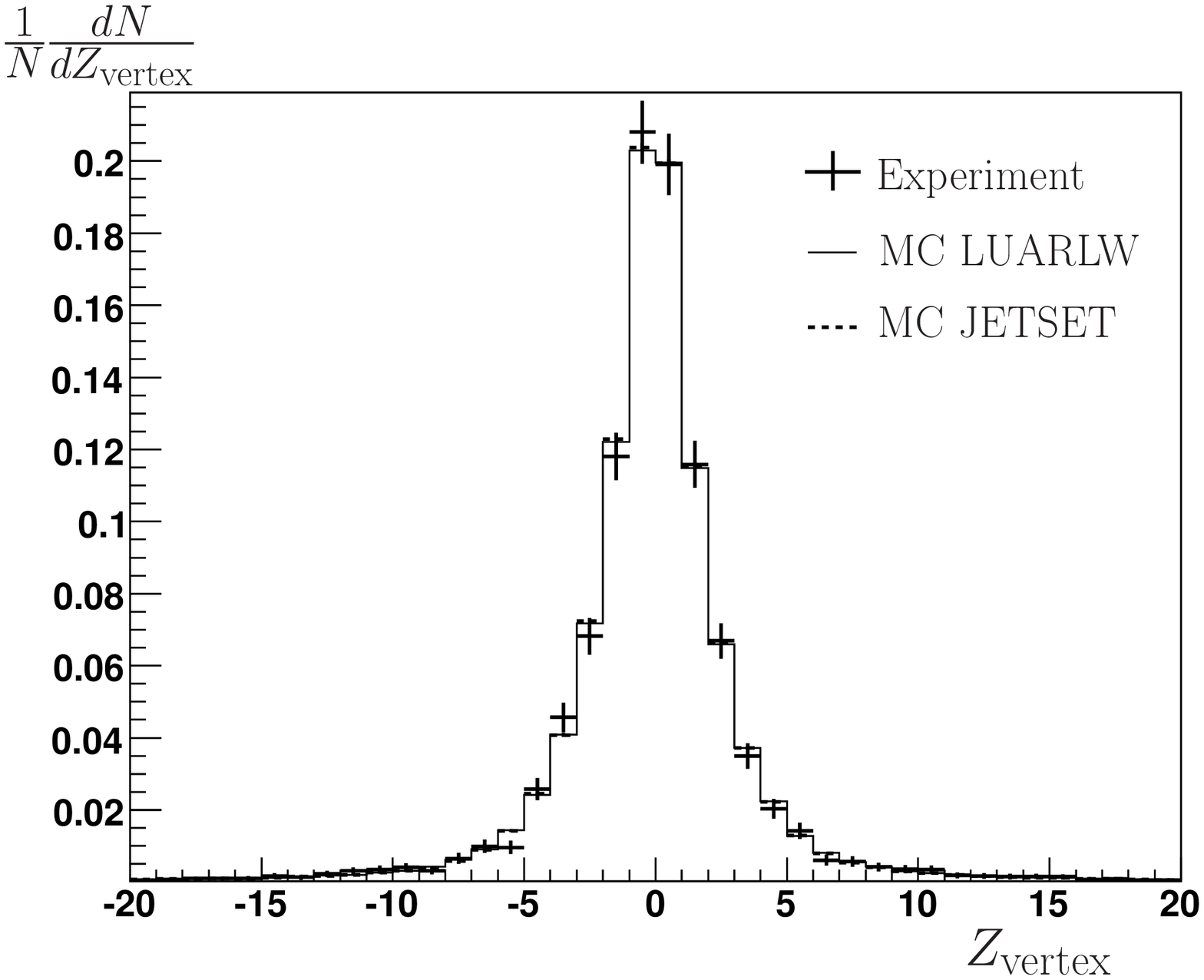}  
\caption{{ 
 Properties of hadronic events produced in $\text{uds}$ continuum at 3.12 GeV.
 Here $N$ is the number of events, $S_{\text{ch}}$ is sphericity,
 $H_2$ and $H_0$ are Fox-Wolfram moments.
 Integrals of all distributions are normalized to unity.
}}
\label{simdist_fig1}
\end{center}
\end{figure*}
\end{center}

The results are presented in Fig.~\ref{simdist_fig1}, where the most important 
event characteristics obtained in the experiment are compared with those 
in simulation. Good agreement is observed.

Bhabha events required for the precise luminosity determination were 
simulated using the BHWIDE generator~\cite{BHWIDEGEN}.
The detection efficiencies for   $\mu^{+}\mu^{-}$ and $\tau^{+}\tau^{-}$ 
events were obtained using the MC generator described in \cite{BERENDS} 
and the KORALB event generator~\cite{KORALB24}, respectively. 

The \PP and \JP decays  were generated with the tuned version of the 
BES generator~\cite{BESGEN} based on the JETSET~7.4 code. The decay tables 
were updated according to the  PDG edition 2010~\cite{PDG:2010}. 
Details of simulation of $\psi(2S)$ hadronic decays  are
discussed in Ref.~\cite{psi2S:2012}.

Simulation reproduces most important event characteristics
of the $J/\psi$ hadronic decays.
That allows us to introduce minor corrections
to the detection efficiency of \JP hadronic decays presented 
in Table~\ref{tab:fits} required in the upper edge of the experiment
energy range.

The two-photon processes $e^{+}e^{-}\to e^{+}e^{-} X$ are simulated with 
the generators described in Refs.~\cite{BERENDS:EEEE,BERENDS:EEMM,KEDR:EEX}.

\subsection{Event selection and detection efficiencies}\protect\label{subsec:mhsel}
Both experimental and simulated events pass the software 
event filter
during the offline analysis. That allows us to reduce systematic inaccuracy
due to trigger instabilities  and uncertainties in the hardware thresholds.
The software  filter recalculates the PT and ST decisions 
with stringent conditions using a digitized response of the detector subsystems.

To suppress the machine background to an acceptable level, the 
following PT conditions were used by OR:
\begin{itemize}\itemsep=-2pt
\item signals from $\ge$ two  non-adjacent scintillation counters\,,
\item signal from the LKr calorimeter\,, 
\item coincidence of the signals from two CsI endcaps.
\end{itemize}
Signals from two particles with the angular separation $\gtrsim\!20^0$
should satisfy numerous ST conditions.

The MC simulation yields the trigger efficiency of about 0.95 for 
continuum $\text{uds}$ production.
\renewcommand{\arraystretch}{1.2}
\begin{table}[ht!]
\caption{ \label{tab:criteria} {  Selection criteria for 
hadronic events which were used by AND.}}
\begin{center}
\begin{tabular}{l|l}                                                    
\specialrule{.14em}{.07em}{.07em}  
\firsthline
Variable                          &   Allowed range \\ \hline   
$N^{\text{IP}}_{\text{track}}$                  &    $\geq 1$\\ \hline        
$E_{\text{obs}}$                         &    $>1.6~\text{GeV}$ \\ \hline
$E_{\gamma}^{\text{max}}/E_{\text{beam}}$       &    $<0.8$ \\ \hline
$E_{\text{cal}}^{\text{tot}}$                   &    $>0.75~\text{GeV}$ \\ \hline
$H_2/H_0$                         &    $<0.85$ \\ \hline
$|P_{\text{z}}^{\text{miss}}/E_{\text{obs}}|$          &    $<0.6$ \\ \hline
$E_{\text{LKr}}/E_{\text{cal}}^{\text{tot}}$           &    $>0.15$ \\ \hline
$|Z_{\text{vertex}}|$                    &    $<20.0~\text{cm}$\\ \hline                                    
 \multicolumn{2}{c}{ $N_{\text{particles}} \geq 4~\text{or}~\tilde{N}^{\text{IP}}_{\text{track}} \geq 2 $}\\          
\hhline{==}
\end{tabular}
\end{center}
\end{table}
\renewcommand{\arraystretch}{1.1}
\begin{table}[t!]
\caption{ \label{tab:def} {Detection efficiency for the $\text{uds}$ continuum} in $\%$ (statistical errors only).}
\begin{center}
\begin{tabular}{cccc}    
\specialrule{.14em}{.07em}{.07em} 
\firsthline
Point               &  $\epsilon_{JETSET}$      &  $\epsilon_{LUARLW}    $   &   $\delta \epsilon/\epsilon$    \\ \hline               
   \multicolumn{4}{c}{ Scan 1}                 \\ \hline
1           & $75.5 \pm 0.1 $   & $75.0 \pm 0.1 $& $-0.7\pm 0.2 $   \\ \hline   
2           & $76.9 \pm 0.1 $   & $76.2 \pm 0.1 $& $-0.9\pm 0.2  $  \\ \hline   
3           & $77.0 \pm 0.1 $   & $77.0 \pm 0.1 $& $ 0.0\pm 0.2  $   \\ \hline  
4           & $78.1 \pm 0.1 $   & $77.4 \pm 0.1 $& $-0.9\pm 0.2  $   \\ \hline  
5           & $78.3 \pm 0.1 $   & $78.2 \pm 0.1 $& $-0.1\pm 0.2  $   \\ \hline  
6           & $79.6 \pm 0.1 $   & $78.6 \pm 0.1 $& $-1.3\pm 0.2  $   \\ \hline  
7           & $80.8 \pm 0.1 $   & $79.2  \pm 0.1$& $-2.0\pm 0.2 $    \\ \hline  
                           \multicolumn{4}{c}{ Scan 2}      \\ \hline 
1          & $75.3  \pm 0.1$     & $74.9\pm 0.1$  & $-0.5\pm 0.2$        \\ \hline 
2          & $75.9 \pm 0.1$      & $75.1\pm 0.1$ &  $-1.1\pm 0.2$        \\ \hline 
3          & $77.5 \pm 0.1$      & $77.3\pm 0.1$ &  $-0.3\pm 0.2$         \\ \hline  
4          & $78.7 \pm 0.1$      & $78.0\pm 0.1$ &  $-0.9\pm 0.2$         \\ \hline  
5          & $78.8 \pm 0.1$      & $78.7\pm 0.1$ &  $-0.1\pm 0.2$         \\ \hline  
6          & $80.0 \pm 0.1$      & $79.0\pm 0.1$ &  $-1.3\pm 0.2$         \\ \hline  
7          & $80.9 \pm 0.1$      & $79.4\pm 0.1$ &  $-1.9\pm 0.2$         \\         
\hhline{====}           
\end{tabular}
\end{center}
\end{table}
\renewcommand{\arraystretch}{1.1}
Selection criteria for multihadron events  are listed in 
Table~\ref{tab:criteria}, 
and their description is provided below. Here $N^{IP}_{track}$ is the number 
of tracks from a common vertex in the interaction 
region defined by: \mbox{$\,\rho\!<\!5$}~mm,\, \mbox{$|z_0|\!<\!130$}~mm, 
where $\rho$ is the track impact parameter relative to the beam axis
and $z_0$--coordinate of the closest approach point.
The $\tilde{N}^{IP}_{track}$ is the number of tracks satisfying the 
conditions above with $E/p$  less than 0.6, where $E/p$ means the ratio of 
the energy deposited in the calorimeter to the measured momentum of the 
charged particle.
The multiplicity $N_{\text{particles}}$ is a sum of the number of charged tracks 
and the number of neutral particles detected in the calorimeters.

The observable energy  $E_{\text{obs}}$ is defined as a sum  
of the photon energies measured in the electromagnetic calorimeter  and
charged particle energies computed from the track momenta assuming pion masses.
The observable energy cut and limitation on the ratio of the energy of 
the most energetic photon 
to the beam energy $E_{\gamma}^{\text{max}}/E_{\text{beam}}$ suppress production 
of hadronic events at low center-of-mass energies 
through initial state radiation  and thus reduce the uncertainty of 
radiative corrections.   
The total calorimeter energy $E_{\text{cal}}^{\text{tot}} $ is defined as a sum 
of the energies of all clusters 
in the electromagnetic calorimeter.  The cut on it suppresses background 
from cosmic rays.
The cut on the ratio of Fox-Wolfram moments $H_{2}/H_{0}$  is efficient 
for suppression of the $\ee\!\to\!\ee\gamma\,$
background, that of cosmic rays and some kinds of the machine background.
The background from two-photon and beam-gas events is suppressed by the cut 
on the ratio 
$|P_{\text{z}}^{\text{miss}}/E_{\text{obs}}|$,  where $P_{\text{z}}^{\text{miss}}$ 
is the $\text{z}$ component of missing momentum .
The background from beam-gas events was also suppressed by the cut 
on the ratio  $E_{\text{LKr}}/E_{\text{cal}}$ 
of the energy deposited in the LKr calorimeter and total calorimeter energy.
The event vertex position  $Z_{\text{vertex}}$ is the weighted average
of the $z_0$'s of
the charged tracks. The cut on the $|Z_{\text{vertex}}|$ suppresses
background due to beam-gas, beam-wall and cosmic rays.

For additional suppression of the background induced by cosmic rays 
a veto from the muon 
system was required in the cases when more than two tracks did not cross 
the interaction region or
the event arrival time determined by TOF relative to the bunch crossing
was less than -7 ns or larger than 12 ns. 

The detection efficiency for hadronic events corresponding
to the selection criteria described above is presented in
Table~\ref{tab:def} for seven data points in which the $R$ ratio
was measured. It was determined using two versions of the event
simulation.

\subsection{Luminosity determination}\protect\label{subsec:lum}
The integrated luminosity at each point was
determined using Bhabha events detected in the LKr calorimeter in the
polar angle range $41^{\circ}\!<\!\theta\!<\!159^{\circ}$. For the
cross check we used Bhabha events in the endcap CsI calorimeter with
$20^{\circ}\!<\!\theta\!<\!32^{\circ}$ and
$148^{\circ}\!<\!\theta\!<\!160^{\circ}$.

The criteria for \ee event selection are listed below:

\begin{itemize}\itemsep=-2pt
\item  two clusters, each with the energy above $20\%$ of the beam 
energy and the angle
       between them exceeding $162^{\circ}$,
\item the total energy of these two clusters exceeds
         the beam energy,
\item the calorimeter energy not associated  with
         these two clusters does not exceed 20$\%$ of the total.
\end{itemize}
The tracking system was used only to reject the background from 
$\ee\!\to\!\gamma\gamma$ and $\ee\!\to\!\text{\it hadrons}$.

\subsection{Physical background}\protect\label{sec:physbackground}

To measure $R$ values, we took into account the physical background 
contributions 
from the QED processes $\ee\to\ee$, $\ee\to\mumu$ and $\ee\to\tau^{+}\tau^{-}$.
The sum of contributions from  $\ee\to\ee$ and $\ee\to\mumu$ 
 production to the observed cross section is less than 0.1~nb.
The uncertainties in the detection efficiency of $\ee\to\ee$ and
 $\ee\to\mumu$ processes introduce about $0.1\%$ uncertainty in the $R$ value.

The contributions of $\tau^{+}\tau^{-}$ production are 
about 0.2~nb and 0.3~nb  at two highest energy points, respectively, 
which induce a systematic uncertainty of less than $0.1\%$
in the $R$ ratio.

The two-photon interactions, which are the main source of background after 
$\tau^+\tau^-$ production,  were studied with a simulation of 
$\ee\to\ee X$~events. We  found that the contribution of 
two-photon events to the continuum cross section grows from $0.2\%$ 
at 3.12 GeV to $0.5\%$ at 3.72 GeV.
The estimated uncertainty in the $R$ value due to this contribution 
varies from $0.1\%$ to $0.2\%$.

\subsection{Correction for residual background}\protect\label{sec:background}
The contribution of residual machine background to the observed cross section 
was estimated using runs with separated $e^{+}$ and $e^{-}$ bunches. 

The residual background was evaluated and subtracted using
 the number of events which passed selection criteria in the background runs
under the assumption that the background rate is proportional to the beam
current and the measured vacuum pressure.
As an alternative, we assumed that the background rate is proportional to
the current only. The difference between the numbers 
of background events obtained with the two assumptions was
considered as an uncertainty estimate  at given energy point.
The background values and their uncertainties at each energy point 
are presented in Table~\ref{tab:background}.
\begin{table}[h!]
\caption{{\label{tab:background} The residual machine 
background in $\%$ of the observed cross section}}
\begin{center}
\begin{tabular}{ccc}
\specialrule{.14em}{.07em}{.07em} 
\firsthline
Point   &   Scan 1&  Scan 2 \\\hline        
    1   &   $1.3\pm0.2\pm0.4$& $1.3\pm 0.2\pm 0.4$ \\ \hline      
    2   &   $2.4\pm0.4\pm0.5$& $2.7\pm 0.4\pm 0.5$ \\ \hline    
    3   &   $2.7\pm0.5\pm0.4$& $3.0\pm 0.5\pm 0.4$ \\ \hline   
    4   &   $2.9\pm0.5\pm0.4$& $3.6\pm 0.6\pm 0.4$ \\ \hline     
    5   &   $3.1\pm0.6\pm0.5$& $3.3\pm 0.5\pm 0.5$ \\ \hline    
    6   &   $2.7\pm0.5\pm0.4$& $3.7\pm 0.6\pm 0.4$ \\ \hline     
    7   &   $2.1\pm0.4\pm0.2$& $2.2\pm 0.3\pm 0.2$ \\          
\hhline{===}   
\end{tabular}
\end{center}
\end{table}

\subsection{Radiative correction}\protect\label{sec:radeff}
The radiative correction factor was calculated according to
 Eq.~\eqref{eq:RadDelta} using the compilation of the vacuum polarization data by the CMD-2 group
 \cite{Actis:2010gg} and the relation between
$R(s)$ and the hadronic part of the vacuum polarization $\Pi_{\text{hadr}}(s)$:
\begin{equation}
R(s)=-\frac{3}{\alpha} \Imag \Pi_{\text{hadr}}(s).
\end{equation}

To calculate the operator $\tilde{\Pi}$ and 
the quantity $\tilde{R}$ for Eq.~\eqref{eq:RadDelta}
we have subtracted  analytically the contribution of the \JP and \PP
from data obtained by the CMD-2 group.

\begin{figure}[!ht]
\includegraphics[width=0.48\textwidth]{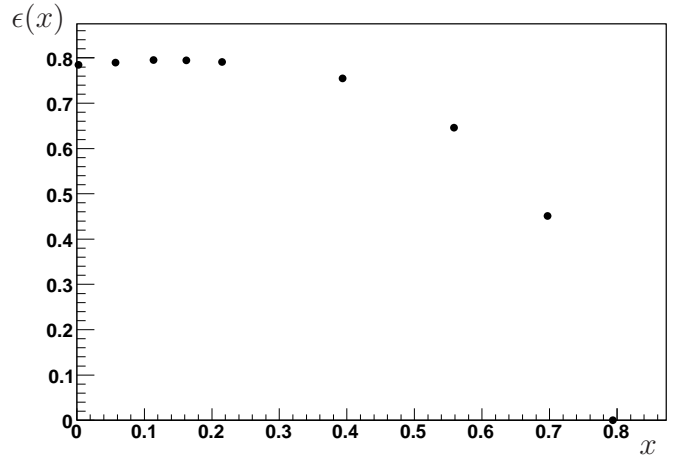}
\caption{{\label{fig:radeff} Hadronic detection efficiency versus variable $x$ of Eq.~\eqref{eq:RadDelta} at  3.52 GeV in the first scan.
}}
\end{figure}

The dependence of the detection
efficiency on the energy radiated in the initial state was simulated
with the $\text{LUARLW}$ generator which allowed us to simulate $\text{uds}$ continuum
below 3.12~GeV.
The $x$ dependence of the detection efficiency  
is shown in Fig.~\ref{fig:radeff}. 

Table \ref{tab:delta} contains values of the radiative correction and their 
systematic uncertainties which are discussed in Sec.~\ref{sec:raderr}.
\begin{table}[th!]
\caption{ \label{tab:delta}{Radiative correction factor $1+\delta$}}
\begin{center}
\begin{tabular}{ccc}    
\specialrule{.14em}{.07em}{.07em}  
\firsthline                                         
 Point              &       Scan 1    &   Scan 2    \\ \hline
1              & $1.0941 \pm 0.0066 $   &$1.1074\pm 0.0066$     \\ \hline
2              & $1.0949 \pm 0.0055 $   &$1.1049\pm 0.0055$     \\ \hline
3              & $1.0959 \pm 0.0055 $   &$1.1100\pm 0.0056$     \\ \hline
4              & $1.0982 \pm 0.0044 $   &$1.1094\pm 0.0044$     \\ \hline
5              & $1.1032 \pm 0.0044 $   &$1.1102\pm 0.0044$    \\ \hline
6              & $1.1021 \pm 0.0044 $   &$1.1098\pm 0.0044$    \\ \hline
7              & $1.1049 \pm 0.0055 $   &$1.1067\pm 0.0055$    \\ 
\hhline{===}   
\end{tabular}
\end{center}
\end{table}

\subsection{$J/\psi$ and $\psi(2S)$ contributions}\protect\label{sec:Jpsi}

To calculate contributions of narrow resonances to the observed
cross section we used the detection efficiencies obtained from the fits. 
The values presented in Table \ref{tab:fits} were corrected for the
presence of ISR photons.
The corrections were obtained via simulation of $J/\psi$ and $\psi(2S)$
hadron decays at each energy point.
These results are presented in Table~\ref{tab:Eff}.
The systematic uncertainties of the $J/\psi$ and $\psi(2S)$ detection
efficiencies are due to the uncertainty in 
the beam energy determination and the detector instability.

\begin{table}[!ht]
\caption{\label{tab:Eff}{Detection efficiency for the $J/\psi$ and $\psi(2S)$ hadronic decays 
of interest and its variation in the experiment energy range.}}
\begin{center}
\begin{tabular}{lcc} 
\specialrule{.14em}{.07em}{.07em} 
\firsthline      
Resonance   &Detection efficiency, $\%$ &  $\Delta\epsilon/\epsilon$, $\%$   \\ \hline         
               \multicolumn{3}{c}{Scan 1}     \\ \hline           
$J/\psi$   & $76.1\pm 1.3 \pm  0.5$  &   $+1.2 \pm 0.1$     \\ \hline    
$\psi(2S)$ & $83.8\pm 2.3 \pm  0.9$  &   $+0.1 \pm 0.1$     \\ \hline                         
               \multicolumn{3}{c}{Scan 2}     \\ \hline           
$J/\psi$   & $75.1\pm 1.4 \pm  0.5$  &    $+1.3 \pm 0.1$     \\ \hline    
$\psi(2S)$ & $83.0\pm 2.0 \pm  0.9 $ &    $+0.1 \pm 0.1$     \\
\hhline{===}   
\end{tabular}
\end{center}
\end{table}

Simulation of $J/\psi$ hadron decays yields the detection 
efficiencies of $0.771\pm0.001$ and $0.767 \pm 0.001$ for two scans, respectively. 
The detection efficiencies obtained from simulation of $\psi(2S)$ 
hadronic decays are  $0.816 \pm 0.001$ and $0.817\pm 0.001$ for two scans,
respectively. 
For both resonances the detection efficiencies obtained by simulation 
agree with the fit results within the estimated errors.

\subsection{Results of energy scans}\protect\label{sec:rscans}
The results of $R$ measurement obtained in energy scans are presented 
in Table  \ref{tab:rvalues_sc}.
\begin{table}[ht!]
\caption{\label{tab:rvalues_sc}{ Resulting $R$ values with their statistical errors for two scans.}}
\begin{center}
\begin{tabular}{lccc}     
\specialrule{.14em}{.07em}{.07em} 
\firsthline      
     Point                & Scan 1 & Scan 2     \\ \hline                
 1              & $2.194 \pm 0.122 $&  $2.239 \pm 0.131$   \\ \hline
 2              & $2.195 \pm 0.078 $&  $2.148 \pm 0.082$   \\ \hline
 3              & $2.233 \pm 0.072 $&  $2.152 \pm 0.089$   \\ \hline
 4              & $2.152 \pm 0.066 $&  $2.190 \pm 0.078$   \\ \hline
 5              & $2.173 \pm 0.062 $&  $2.247 \pm 0.086$   \\ \hline
 6              & $2.209 \pm 0.110 $&  $2.198 \pm 0.070$   \\ \hline 
 7              & $2.195 \pm 0.116 $&  $2.183 \pm 0.084$   \\  
\hhline{====}
\end{tabular}
\end{center}
\end{table}

\section{Systematic uncertainties and results}\protect\label{sec:systerr}

\subsection{Systematic uncertainty of absolute luminosity determination}\protect\label{sec:lumerr}
The major contributions to the uncertainty of the
absolute luminosity determination with the LKr calorimeter are 
presented in Table~\ref{tab:lumerr}.

\begin{table}[h]
\caption{{ \label{tab:lumerr} Systematic uncertainties of
the luminosity determination.}}
\begin{center}
\begin{tabular}{lc}
\specialrule{.14em}{.07em}{.07em} 
\firsthline      
Source &  Uncertainty, $\%$   \\ \hline                            
Calorimeter response          & 0.7  \\  \hline
Calorimeter alignment         & 0.2  \\  \hline
Polar angle resolution        & 0.2  \\  \hline
Cross section calculation     & 0.5  \\  \hline 
Background                    & 0.1  \\  \hline
MC statistics                 & 0.1 \\   \hline
Variation of cuts             & 0.6  \\  
\specialrule{.14em}{.07em}{.07em}  
Sum in quadrature      & 1.1  \\
\hhline{==}
\end{tabular}
\end{center}
\end{table}

The uncertainty due to the imperfect simulation of the calorimeter
response was estimated by variation of relevant simulation parameters such as
the accuracy of the electronic channel calibration, the geometrical
factor controlling sensitivity to the energy loss fluctuations between
calorimeter electrodes etc.

The LKr calorimeter was aligned with respect to the drift chamber using
cosmic tracks reconstructed in the DC.
The interaction point position and direction of the beam line were determined
using the primary vertex distribution of multihadron events. 
The luminosity   uncertainty  due to inaccuracy of the alignment  is less 
than $0.2\%$.

The difference in the polar angle resolutions observed in  experiment
and predicted by simulation causes an uncertainty in the luminosity 
measurement, because events migrate  into or out of the fiducial volume.

The uncertainty of the theoretical Bhabha cross section
was estimated comparing the results obtained with
the BHWIDE~\cite{BHWIDEGEN} and MCGPJ~\cite{MCGPJ} event generators.
It agrees with the errors quoted by the authors.

The background to the Bhabha process  from the 
$J/\psi$ and $\psi(2S)$ decays and reactions 
$\ee \to \mu\mu(\gamma)$ and $\ee \to \gamma\gamma $ 
was estimated using MC simulation. 
It contributes less than 0.2\% to the observed $e^{+}e^{-}$ cross section
at seven energy points presented in Table~\ref{tab:epoints}.
At the auxiliary points of the scan serving for the determination 
of the $J/\psi$ and $\psi(2S)$ signal
magnitude the contributions of the resonance decays to $e^{+}e^{-}$  
were accounted for in the fits.
We also considered a contribution of residual machine background  
to Bhabha events which is about $0.1\%$.
The residual luminosity uncertainty due to background does not exceed $0.1\%$. 

To evaluate the effect of other possible sources of systematic uncertainty,
the variation of the cuts was performed within the fiducial
region in which good agreement between the MC simulation and 
experiment is observed.

Differences of integrated luminosities obtained using the LKr and CsI 
calorimeters in two scans are $0.5 \pm 0.5 \%$ and $0.0\pm 0.5\%$, respectively.
That is consistent with the estimates in Table~\ref{tab:lumerr}.

\subsection{Uncertainty due to imperfect simulation of continuum}\protect\label{sec:mchadrerr}
The imperfect simulation of $\text{uds}$ continuum contributes
significantly to the systematic uncertainty in  $R$. 
Considering the detection efficiencies 
reported in Table~\ref{tab:def} 
 obtained with the JETSET
  and LUARLW hadronic generators one can evaluate the 
systematic uncertainty related to the detection efficiency. 
The maximal deviation of $1.3\%$ is taken as the systematic 
uncertainty for the energy range 3.12-3.62 GeV.
Our estimate of the systematic uncertainty due to the $\text{uds}$ continuum 
generator is more conservative than the value $0.5\%$ used in 
Ref.~\cite{BES:R2009} with the LUARLW generator in this energy range.
At the energy of 3.72 GeV our estimation of this uncertainty is $2\%$.

There is a systematic uncertainty in the observed multiplicity 
related to the track reconstruction efficiency, which
is not exactly the same for the experimental data and
simulation. The multiplicity together with other event parameters
was employed for the JETSET parameter tuning limiting the tuning accuracy.
The reconstruction efficiency was studied using Bhabha events and low-momentum cosmic tracks and the
appropriate correction was introduced in the MC simulation.
The uncertainty of the correction introduces the additional systematic
uncertainty of about 0.5\%.

The contributions to the detection efficiency uncertainty due to 
imperfect simulation of $\text{uds}$ continuum 
are summarized in Table~\ref{tab:mcerr}.

\begin{table}[ht]
\caption{{\label{tab:mcerr} Systematic uncertainties of the
detection efficiency due to $\text{uds}$ continuum simulation. 
}}
\begin{center}
\begin{tabular}{lcc}                     
\specialrule{.14em}{.07em}{.07em} 
\firsthline      
Source                & \multicolumn{2}{c}{ Uncertainty, $\%$}  \\ \hline
                      & Points 1-6  & Point 7  \\ \hline
$\text{uds}$ simulation      &  1.3        & 2.0    \\ \hline
Track reconstruction  &  0.5        & 0.5  \\ \hline
MC statistics         &  0.2        & 0.2   \\ 
\specialrule{.14em}{.07em}{.07em} 
Sum in quadrature     &  1.4        & 2.1 \\
\hhline{===}
\end{tabular}
\end{center}
\end{table}

\subsection{Systematic uncertainty of the radiative correction}\protect\label{sec:raderr}
The main sources of systematic uncertainty in the radiative correction factor 
at each energy point are listed in Table \ref{tab:raderr}.
The four contributions were evaluated and summed up in quadrature.

To estimate the uncertainty related to a choice of the vacuum
polarization operator approximation, that from CMD-2~\cite{Actis:2010gg} was 
replaced with the approximation employed in
the BES generator~\cite{BESGEN}. The variation reaches 0.4\% at the
points closest to $J/\psi$ and $\psi(2S)$ and drops down to $0.1\%$
at the other points.
The contribution denoted as $\delta R(s)$ is
related to the $R(s)$ uncertainty.
It is less than 0.5\% for the entire  energy range.
The contribution $\delta \epsilon(s)$  of about 0.2\% is related 
to the uncertainty in the $\epsilon(s)$ dependence.
A calculation of the radiative corrections according to
Eq.~\eqref{eq:RadDelta} requires
the interpolation of the detection efficiency presented in
Fig.~\ref{fig:radeff} as a
function of $x$. The contribution
$\delta_{\rm calc}$ is due to the relatively small number of points
where the efficiency was calculated by Monte Carlo. 
It was estimated comparing the results obtained using the linear
interpolation and the quadratic one.

\begin{table}[h!]
\caption{{ \label{tab:raderr} Systematic uncertainties of the 
radiative correction.}}
{\small
\begin{center}
\begin{tabular}{cccccc}
\specialrule{.14em}{.07em}{.07em} 
\firsthline 
             &  \multicolumn{5}{c}{ Uncertainty, $\%$ }     \\ \hline
  Point  &  \multicolumn{4}{c}{Contributions} & Total \\ \hline
   & $\Pi$ approx. &  $\delta R(s)$  &$\delta \epsilon(s)$ & $\delta_{calc}$ &  \\ \hline    
    1  &   0.3 & 0.5&0.2&0.2 &0.6     \\ \hline       
    2  &   0.1 & 0.4&0.2&0.2 &0.5    \\ \hline    
    3  &   0.1 & 0.4&0.2&0.2 &0.5   \\ \hline 
    4  &   0.1 & 0.3&0.2&0.2 &0.4    \\ \hline
    5  &   0.1 & 0.3&0.2&0.2 &0.4    \\ \hline  
    6  &   0.1 & 0.3&0.2&0.2 &0.4    \\ \hline 
    7  &   0.4 & 0.3&0.2&0.2 &0.5    \\ 
\hhline{======}
\end{tabular}
\end{center}
}
\end{table}

\subsection{Detector-related uncertainties in $R$}\label{err:deterr}
The systematic uncertainties related to the efficiency of the track 
reconstruction were considered in Sec.~\ref{sec:mchadrerr}.

The main source of the trigger efficiency uncertainty is 
that of the calorimeter thresholds in the secondary trigger.
The estimate of about $0.2\%$ was obtained varying the threshold
in the software event filter. 

The trigger efficiency  and the event selection efficiency depend 
on the calorimeter response to hadrons. 
The uncertainty related to the simulation of nuclear interaction
was estimated by comparison of the efficiencies obtained with the
packages GHEISHA~\cite{Fesefeldt:1985yw} and FLUKA~\cite{Fasso:2005zz}
which are implemented in GEANT~3.21~\cite{GEANT:Tool}. 
The relative difference was about 0.2\%.

The effect of other possible sources of the detector-related uncertainty 
was evaluated
by varying the event selection cuts that are presented in 
Table~\ref{tab:criteriavar}. 
All variations of $R$ observed were  smaller than their statistical 
errors and can originate from 
the already considered sources of uncertainties or the statistical 
fluctuations, nevertheless we included 
them in the total uncertainty to obtain conservative error estimates.

\begin{table}[h!]
\caption{\label{tab:criteriavar} { $R$ uncertainty due to 
variation of the selection criteria for hadronic events.}}
\begin{center}
\begin{tabular}{llc}      
\specialrule{.14em}{.07em}{.07em} 
\firsthline 
Variable                                       &   Range variation & $R$ variation in \% \\ \hline   
$E_{\text{obs}}$                               &    $> 1.4 \div 1.8~\text{GeV}$ & 0.3 \\ \hline
$E_{\gamma}^{\text{max}}/E_{\text{beam}}$      &    $< 0.6 \div  0.9 $ & 0.3 \\ \hline
$E_{\text{cal}}^{\text{tot}}$                  &    $> 0.5 \div  0.75~\text{GeV}$ & 0.2 \\ \hline
$H_2/H_0$                                      &    $< 0.75 \div 0.9$ &  0.3\\ \hline
$|P_{\text{z}}^{\text{miss}}/E_{\text{obs}}|$  &    $< 0.6 \div 0.8$ & 0.2     \\ \hline
$E_{\text{LKr}}/E_{\text{cal}}^{\text{tot}}$   &    $> 0.15 \div0.25$& 0.1\\ \hline
$|Z_{\text{vertex}}|$                          &    $< 20.0\div15.0~\text{cm}$&  0.2  \\ 
\specialrule{.14em}{.07em}{.07em}  
\multicolumn{2}{c}{Sum in quadrature} &      0.6   \\                             
\hhline{===}
\end{tabular}
\end{center}
\end{table}

\subsection{Summary of systematic uncertainties and results}\protect\label{sec:errsummary}
The major sources of the systematic uncertainty on the $R_{\text{uds}}$ value 
are listed in Table~\ref{tab:rerr}.
\renewcommand{\arraystretch}{1.}
\begin{table*}[th!]
\caption{{ \label{tab:rerr} $R_{\text{uds}}$ systematic uncertainties (in $\%$) 
assigned to each energy point.}}
\begin{center}
\begin{tabular}{cccccccc}
\specialrule{.14em}{.07em}{.07em} 
\firsthline 
                                   & Point 1  & Point 2  &   Point 3  & Point 4  &   Point 5 & Point 6  & Point 7          \\ \hline    
Luminosity                 &   1.1      &  1.1    &     1.1       & 1.1        &   1.1      & 1.1    & 1.1          \\ \hline     
Radiative correction       &   0.6      &  0.5    &     0.5       & 0.4        &   0.4      & 0.4    & 0.5        \\ \hline    
Continuum simulation       &   1.4      &  1.4    &     1.4       & 1.4        &   1.4      & 1.4    & 2.1         \\ \hline    
$e^+e^-X$  contribution    &   0.1      &  0.1    &     0.1       & 0.2        &   0.2      & 0.2    & 0.2          \\ \hline     
$l^+l^-$  contribution     &   0.1      &  0.1    &     0.1       & 0.1        &   0.1      & 0.2    & 0.2          \\ \hline     
Trigger efficiency         &   0.2      &  0.2    &     0.2       & 0.2        &   0.2      & 0.2    & 0.2          \\ \hline    
Nuclear interaction        &   0.2      &  0.2    &     0.2       & 0.2        &   0.2      & 0.2    & 0.2          \\ \hline                  
Cuts variation             &   0.6      &  0.6    &     0.6       & 0.6        &   0.6      & 0.6    & 0.6          \\ \hline    
                                                   \multicolumn{8}{c}{\emph{Scan 1}} \\\hline    
$J/\psi$ and  $\psi(2S)$ contribution     &   2.7      &  0.5  &     0.3    & 0.2    &   0.2   & 0.1    & 1.4        \\ \hline    
Machine background                        &   1.1      &  0.8  &     0.7    & 0.7    &   0.9   & 0.7    & 0.7         \\ 
\specialrule{.1em}{.05em}{.05em} 

Sum in quadrature                         &   3.5      &  2.2  &     2.1    & 2.1    &   2.2   & 2.1    & 3.0        \\ \hline 
                                                    \multicolumn{8}{c}{ \emph{Scan 2} } \\ \hline   
$J/\psi$ and  $\psi(2S)$ contribution     &   2.8     &  0.6    &    0.3    & 0.2    &   0.2   & 0.1      & 1.3         \\ \hline      
Machine background         &   1.1     &  0.8       &    0.7      & 0.8       &   0.8      & 0.7          & 0.5         \\ 
\specialrule{.1em}{.05em}{.05em} 
Sum in quadrature           &   3.6     &  2.2       &    2.1      & 2.1       &   2.1     &   2.1        &  2.9         \\ 
\specialrule{.1em}{.05em}{.05em} 
Correlated in two scans     &   2.3    &   1.9       &    1.8     & 1.8     &    1.8     &   1.8        &  2.5             \\ 
\hhline{========}
\end{tabular}
\end{center}
\end{table*}

At each energy point we divide the systematic uncertainty into a 
common uncertainty that is correlated
in two scans for given energy and uncorrelated uncertainty that is 
independent for each scan.
During data collection at given energy point the relative 
beam energy variation was less than $10^{-3}$  allowing us to
neglect this source of uncertainty.

As mentioned above, the contribution of narrow resonances 
to $R(s)$ is not negligible in the resonance region.
This contribution was found  analytically using "bare"  parameters 
of the resonances, 
which were calculated based on the PDG data~\cite{PDG:2014}.

The results of the two scans were weighted using 
their statistical uncertainties and the uncorrelated parts of the systematic
ones. The formal description of the weighting procedure can be found in Ref.~\cite{psi2S:2012}.
The obtained $R_{\text{uds}}$ and $R$ values as well as luminosity-weighted 
average center-of-mass 
energies are presented in Table~\ref{tab:rvalues}. 

\begin{table}[ht!]
\caption{\label{tab:rvalues}{ Measured values of $R_{\text{uds}}(s)$ and $R(s)$
with statistical and systematic uncertainties.}}
\begin{center}
\begin{tabular}{lc}      
\specialrule{.14em}{.07em}{.07em} 
\firsthline 
     $\sqrt{s}$, MeV  &               $R_{\text{uds}}(s)~\{ R(s)\}$    \\ \hline                
 $3119.9  \pm 0.2$  & $2.215\{ 2.237\} \pm 0.089 \pm 0.066$ \\ \hline%
 $3223.0  \pm 0.6$  & $2.172\{ 2.173\} \pm 0.057 \pm 0.045$    \\ \hline%
 $3314.7  \pm 0.7$  & $2.200\{ 2.200\} \pm 0.056 \pm 0.043$   \\ \hline%
 $3418.2  \pm 0.2$  & $2.168\{ 2.168\} \pm 0.050 \pm 0.042$   \\ \hline%
 $3520.8  \pm 0.4$  & $2.200\{ 2.201\} \pm 0.050 \pm 0.044$  \\ \hline%
 $3618.2  \pm 1.0$  & $2.201\{ 2.207\} \pm 0.059 \pm 0.044$ \\ \hline %
 $3719.4  \pm 0.7$  & $2.187\{ 2.211\} \pm 0.068 \pm 0.060$  \\  
\hhline{==}
\end{tabular}
\end{center}
\end{table}

The inaccuracy of $R$ associated with the resonance parameters is negligible in comparison with the others uncertainties, 
so the errors for the values of $R$ and $R_{\text{uds}}$ are the same.

\section{Summary}
We have measured the $R$ and  $R_{\text{uds}}$ values  at seven center-of-mass energies 
between 3.12 and 3.72 GeV.
At most of the energy points the achieved accuracy is about or better 
than  $3.3\%$ at the systematic uncertainty $2.1\%$.
The $R$ values are consistent within errors with the BES 
results~\cite{BES:R2009} and provide more detailed information on the $R(s)$ quantity in this energy 
range.

\begin{figure}[!ht]
\begin{center}
\includegraphics[width=0.44\textwidth]{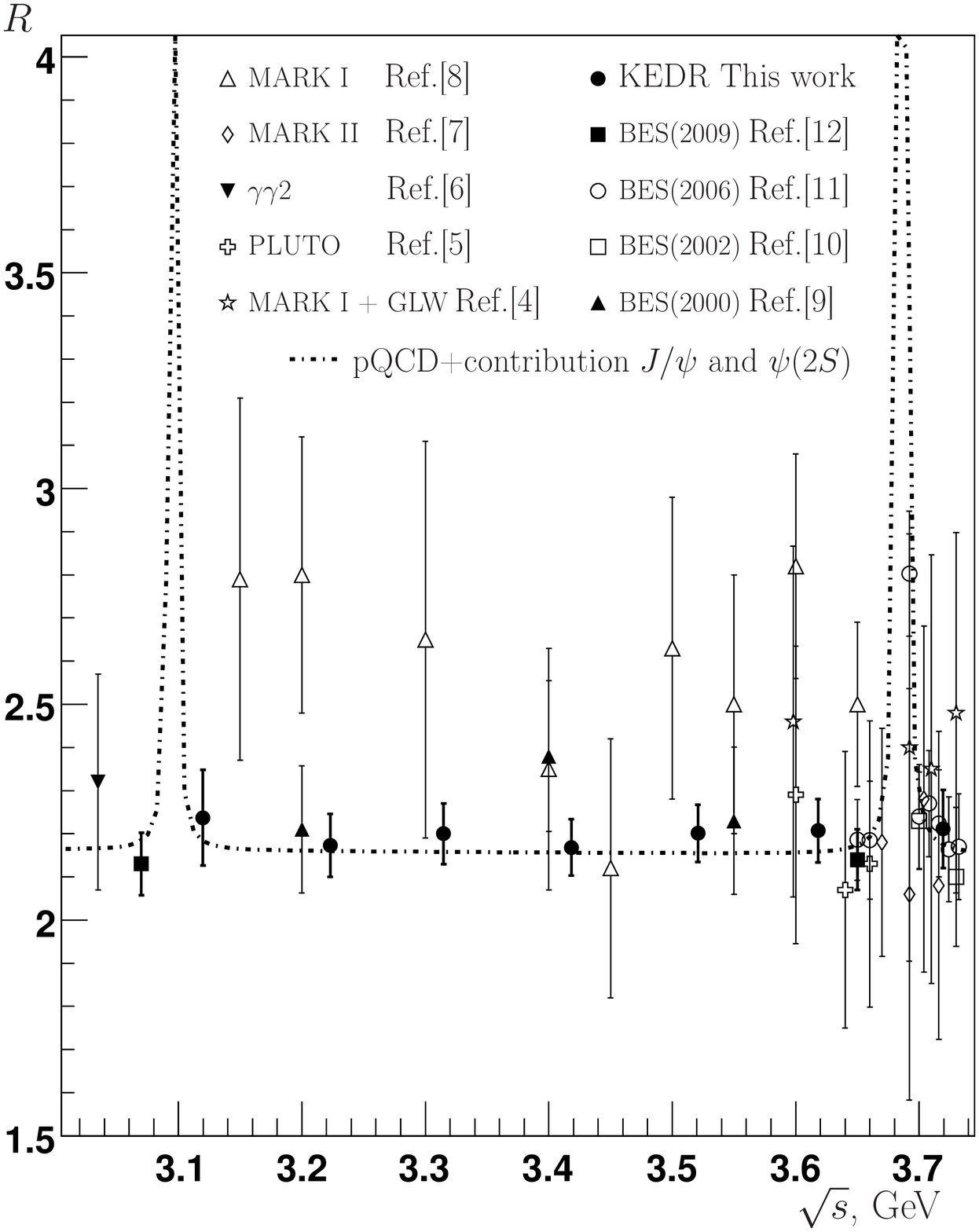}
\end{center}
\caption{{\label{fig:rfinal} The quantity R versus the c.m. energy and 
the sum of the prediction of perturbative QCD and a contribution of narrow resonances.
}}
\end{figure}

The weighted average $\overline{R}_{\text{uds}} = 2.189 \pm 0.022 \pm 0.042$ agrees well 
with $R = 2.16 \pm 0.01$ calculated  according to the pQCD 
expansion~\cite{Baikov:pQCD}  for  $\alpha_{s}(m_{\tau})=0.333\pm0.013$ 
obtained from hadronic $\tau$ decays \cite{Brambilla:2014}. 
The results are shown in Fig.~\ref{fig:rfinal}. 

It is worth noting that while calculating the dispersion integrals in this energy range it is preferable to use the measured $R_{\text{uds}}(s)$ values 
adding the contribution of narrow resonances calculated analytically. Using
the full $R$ values in this case leads to some double counting.

\section*{Acknowledgments}
We greatly appreciate the efforts of the staff of VEPP-4M to provide
good operation of the complex during long term experiments.
The authors are grateful to V.~P.~Druzhinin for useful discussions.

This work has been supported by Russian Science Foundation (project N 14-50-00080).

\end{document}